\documentclass[10pt]{iopart}

\begin{document}

\def\gat{\gamma_a^T}
\def\la{\lambda}
\def\om{\omega}
\def\La{\Lambda}
\def\Th{\theta}
\def\chs{\chi^\star}
\def\Dirac{i\partial\!\!\!\!/}
\def\dirac{i\partial\!\!\!\!/-A\!\!\!\!/}
\def\pip{\Pi_+}
\def\pim{\Pi_-}
\def\pipl{\Pi_+^\star}
\def\pimi{\Pi_-^\star}
\def\gf{\tilde{\gamma}}
\def\rand{\left|_{\partial M}\right. }
\def\ch{\cosh \Theta}
\def\sh{\sinh\Theta}
\def\nen{\Lambda \ch -2\om\sh}
\def\tr{\mbox{Tr}}

\newcommand{\gc}{\gamma _M^0}
\newcommand{\gu}{\gamma _M^1}
\newcommand{\gt}{\tilde{\gamma}_M}
\newcommand{\gce}{\gamma _0}
\newcommand{\gue}{\gamma _1}
\newcommand{\beq}{\begin{eqnarray}}
\newcommand{\eeq}{\end{eqnarray}}
\newcommand{\nn}{\nonumber}

\article[Dirac operator with bag boundary conditions]{\fl Seventh International
Workshop Quantum Field Theory under the Influence of External
Conditions, QFEXT'05, Barcelona, Spain}{Finite temperature properties of the Dirac operator with bag boundary conditions}

\author{C G Beneventano\footnote{Member of CONICET} and
E M Santangelo\footnote{Member of CONICET}}
\address{Departamento de F{\'\i}sica, Universidad Nacional de La Plata
\\Instituto de F{\'\i}sica de La Plata, UNLP-CONICET\\
C.C. 67, 1900 La Plata, Argentina}
\ead{\mailto{gabriela@obelix.fisica.unlp.edu.ar},
\mailto{mariel@obelix.fisica.unlp.edu.ar}}
\begin{abstract}

We study the finite temperature free energy and fermion number for
Dirac fields in a one-dimensional spatial segment, under local boundary conditions, compatible with
the presence of a spectral asymmetry. We discuss in detail the
contribution of this part of the spectrum to the determinant. We evaluate the finite temperature properties of the theory for arbitrary values of the chemical potential.\\[.3cm]
{\bf Subject Classification} 

\end{abstract}
\section{Introduction}\label{intr}

When the Euclidean Dirac operator is considered on
even-dimensional compact manifolds with boundary, its domain can
be determined through a family of local boundary conditions which
define a self-adjoint boundary problem \cite{wipf95-443-201} (the
particular case of two-dimensional manifolds was first studied in
\cite{hras84-245-118}). The whole family is characterized by a
real parameter $\Th$, which can be interpreted as an analytic
continuation of the well known $\Th$ parameter in gauge theories. These boundary conditions can be considered to be the natural
counterpart in Euclidean space of the well known chiral bag
boundary conditions.

One salient characteristic of these local boundary conditions is
the generation of an asymmetry in the spectrum of the Dirac
operator. For the particular case of two-dimensional product
manifolds, such asymmetry was shown, in \cite{bene02-35-9343}, to
be determined by the asymmetry of the boundary spectrum. For other recent work on chiral bag boundary conditions, see \cite{chiral1,chiral2,chiral3}.

In a previous paper \cite{nosbag}, we studied a theory of Dirac fields in one spatial dimension and evaluated its finite-temperature properties for two particular values of the parameter $\Th$, and for restricted ranges of the chemical potential $\mu$, since our aim was to (partially) answer the question posed in \cite{dunne}, as
to whether the fermion number is modified by temperature in low
dimensional bags. Here, we generalize such results to other values of $\mu$.
Such generalization could be of interest in the study of boundary effects in low-dimesional condensed matter systems and also in the treatment of open string theories with a nontrivial twist (analytic extension of $\mu$) of the world sheet \cite{lugo,polchinski}.

In section \ref{Sect2}, we determine the spectrum of the Euclidean
Dirac operator at finite temperature for $\Th=0$. With this spectrum at hand we perform, in section
\ref{Sect3}, the calculation of the partition function via zeta function
regularization for different ranges of the chemical potential.

Section \ref{Sect4} is devoted to the evaluation of the free
energy and the mean fermion number, both at finite and zero temperature.

\section{Spectrum of the Euclidean Dirac operator}\label{Sect2}

In order to study the effect of temperature, we consider a two-dimensional Euclidean space, with the metric (+,+). We take the Euclidean
gamma matrices to be $\gce=\sigma_1$, $\gue=\sigma_2$.
Thus, the Euclidean action is \beq S_E=\int d^2x\,
\bar{\Psi}(\dirac) \Psi\,.\eeq 

The partition function is given by \beq \log{Z}=\log\,
det(\dirac) _{BC}\,.\label{part}\eeq

Here, $BC$ stands for antiperiodic boundary conditions in the
``time" direction ($0\leq x_0 \leq \beta$, with
$\beta=\frac{1}{T}$) and, in the ``space" direction ($0\leq x_1
\leq L$), \beq
 \nn \left.\frac12 (1+\gce ) \Psi \right\rfloor_{0} &=& 0\\
 \left.\frac12 (1+ \gce) \Psi \right\rfloor_{L} &=& 0\,.\label{bc}\eeq

We will follow \cite{actor} in introducing the chemical potential
as an imaginary $A_0=-i\mu$.
 
 In order to evaluate the partition function in the zeta
 regularization  approach, we first determine the eigenfunctions,
 and the corresponding eigenvalues, of the Dirac operator
\beq(\Dirac +i\gce \mu)\Psi=\omega \Psi\,.\eeq

  To satisfy antiperiodic boundary conditions in the
  $x_0$ direction, we expand
  \beq
  \Psi(x_0,x_1)=\sum_{\lambda}e^{i\lambda x_0}\psi(x_1)\,,\eeq
  with
  \beq
  \lambda_l=(2l+1)\frac{\pi}{\beta},\qquad
  l=-\infty,...,\infty\,.\label{lambda}\eeq

  After doing so, and writing $\psi(x_1)=\left(\begin{array}{c}
    \varphi(x_1) \\
    \chi(x_1) \\
  \end{array}\right)$ we have, for each
  $l$,
  \beq \nn (-\tilde{\lambda}_l+\partial_1)\chi&=&\omega
\varphi \\(-\tilde{\lambda}_l-\partial_1)\varphi&=&\omega \chi
\,.\label{diffeq}\eeq

where 
 \beq
\tilde{\lambda}_l=\lambda_l -i\mu= (2l+1)\frac{\pi}{\beta}-i\mu\,.\eeq
  
   It is easy to see that, with the boundary condition in equation (\ref{bc}), no zero
 mode appears. For $\omega \neq 0$ one has, from (\ref{diffeq}),
 \beq \nn &&\partial^2_1\varphi=-\kappa^2\varphi
 \\&&\chi=-\frac{1}{\omega}(\tilde{\lambda}_l+\partial_1)\varphi\,,\eeq
where $\kappa^2=\omega^2-\tilde{\lambda}_l^2$.

For $\kappa \neq 0$, one has for the eigenvalues \beq \omega_{n,l}
=\pm \sqrt{\left(\frac{n\pi}{L}\right)^2+{\tilde{\lambda}_l}^2} \,,\quad
{\rm with}\quad n=1,...,\infty\,,\quad l=-\infty,...,\infty\,.\label{symspec}\eeq This part of
the spectrum is symmetric. In the case $\kappa=0$ one has a set of
$x_1$-independent eigenfunctions, corresponding to
\beq\omega_l=\tilde{\lambda}_l\,.\label{asymspec}\eeq

It is to be noted that $\omega_l=-\tilde{\lambda}_l$ are not eigenvalues.

\section{Partition function}\label{Sect3}

In order to obtain the partition function, as defined in equation (\ref{part}), we must consider the contributions to $\log{{\cal Z}}$ coming from both types of eigenvalues (equations (\ref{symspec}) and (\ref{asymspec}))
\beq \Delta_1=\left. -\frac{d}{ds}\right\rfloor_{s=0}{\zeta}_1
(s)\,,\label{del1mu}\eeq where \beq \fl {\zeta}_1
(s)=(1+(-1)^{-s})\sum_{\begin{array}{c}
  n=1 \\
  l=-\infty \\
\end{array}}^{\infty}\left[{\left(\frac{n\pi}{\alpha L}\right)}^2+
{\left((2l+1)\frac{\pi}{\alpha
\beta}-i\frac{\mu}{\alpha}\right)}^2\right]^{-\frac{s}{2}}\,,\label{z1mu}\eeq
and \beq \Delta_2=\left. -\frac{d}{ds}\right\rfloor_{s=0}{\zeta}_2
(s)\,,\label{del2mu}\eeq where \beq  {\zeta}_2 (s)=\sum_{l=-\infty
}^{\infty}\left[ (2l+1)\frac{\pi}{\alpha
\beta}-i\frac{\mu}{\alpha}\right]^{-s}\,.\label{z2mu}\eeq

As usual, $\alpha$ is a parameter with
dimensions of mass, introduced to render the zeta function
dimensionless.

The analytic extension of $\zeta_2$ requires a careful selection of the cut in the $\omega$-plane (for details, see reference \cite{nosbag}). The result is \beq \fl \nn
{\zeta}_2 (s)&=&
 \left(\frac{2\pi}{\alpha\beta
}\right)^{-s}\left[ \zeta_H \left(s,\frac12-\frac{i\mu
\beta}{2\pi}\right)+\sum_{l=0 }^{\infty}\left[
-(l+\frac12)-i\frac{\mu\beta}{2\pi}\right]^{-s}\right]\\\fl &=&\left(\frac{2\pi}{\beta
\alpha}\right)^{-s}\left[\zeta_H \left(s,\frac12-\frac{i\mu
\beta}{2\pi}\right)+e^{i\pi sign(\mu)s}\zeta_H
\left(s,\frac12+\frac{i\mu \beta}{2\pi}\right)\right]
\label{ext2mu}\,,\eeq
where $\zeta_H(s,x)$ is the Hurwitz zeta function.

The analytic extension of $\zeta_1$ leads to a separate consideration of different $\mu$-ranges, determined by the energies of the zero-temperature problem \cite{nosbag}. We will perform the extension in two of these ranges. The generalization to other ranges will become evident from these two cases. 

\subsection{$ |\mu| < \frac{\pi}{L}$}

This is the range already studied by us. For details, see reference \cite{nosbag}. In this range, (\ref{z1mu}) can be written in terms of its Mellin
transform as \beq
\fl {\zeta}_1(s)=\frac{(1+(-1)^{-s})}{\Gamma(\frac{s}{2})}\int_0^{\infty}dt\,
t^{\frac{s}{2}-1}\!\!\!\!\sum_{\begin{array}{c}
  n=1 \\
  l=-\infty \\\end{array}}^{\infty}\!\!\!\!e^{-t\left[{\left(\frac{n\pi}{\alpha L}\right)}^2+
{\left((2l+1)\frac{\pi}{\alpha
\beta}-i\frac{\mu}{\alpha}\right)}^2\right]}\,.\eeq

This can also be written as \beq \nn {\zeta}_1
(s)&=&\frac{(1+(-1)^{-s})}{{(\sqrt{\pi}})^{s}\Gamma(\frac{s}{2})}\sum_{n=1}^{\infty}\int_0^{\infty}dt\,
t^{\frac{s}{2}-1}e^{-t\pi \left[\left(\frac{n}{\alpha
L}\right)^2+\left(\frac{1}{\alpha \beta}-\frac{i\mu}{\alpha
\pi}\right)^2\right]}\times\\&&\Theta_3\left(\frac{-2t}{\alpha
\beta}\left(\frac{1}{\alpha \beta}-\frac{i\mu}{\alpha\pi
}\right),\frac{4t}{(\alpha \beta)^2}\right)\,,\eeq
where we have used the definition of the Jacobi theta function \hfil\break $\Theta_3
(z,x)=\sum_{l=-\infty}^{\infty}e^{-\pi x l^2} e^{2\pi z l}$.

To proceed, we use the inversion formula for the Jacobi function, \hfil\break $ \Theta_3(z,x)=\frac{1}{\sqrt{x}}e^{(\frac{\pi
z^2}{x})}\Theta_3\left(\frac{z}{ix},\frac{1}{x}\right)$, and
perform the integration over $t$, thus getting 
 \beq \nn {\zeta}_1
(s)&&=\frac{(1+(-1)^{-s})\beta}{2{\alpha}^{-s}{(\sqrt{\pi}})^{s}\Gamma(\frac{s}{2})}
\left[\Gamma\left(\frac{s-1}{2}\right)\frac{{\pi}^{\frac{1-s}{2}}}{
L^{1-s}}\zeta_R (s-1)+\right.\\&&\left.4\left(\frac{\beta L
}{2}\right)^{\frac{s-1}{2}}\!\!\!\sum_{n,l=1}^{\infty}(-1)^l
\left(\frac{l}{n}\right)^{\frac{s-1}{2}}\cosh{(\mu \beta
l)}K_{\frac{s-1}{2}}\left(\frac{nl\pi \beta}{L}\right)\right]\,,
\label{ext1mu}\eeq
where $\zeta_R$ is the Riemann zeta function.

From (\ref{ext1mu}) and (\ref{ext2mu}) both contributions to
$\log{{\cal Z}}$ in this range of $\mu$ can be obtained. They are given by \beq
\Delta_1=-\frac{\beta
\pi}{12L}+\sum_{n=1}^{\infty}\log{(1+e^{-\frac{2n\pi
\beta}{L}}+2\cosh{(\mu \beta)}e^{-\frac{n\pi \beta}{L}})}\,\eeq
and \beq\fl \nn \Delta_2 &=&-\left[\zeta_H^{\prime}\left(0, \frac12
-\frac{i\mu \beta}{2\pi}\right)+\zeta_H^{\prime}\left(0, \frac12
+\frac{i\mu \beta}{2\pi}\right)+i\pi sign(\mu)\zeta_H \left(0,
\frac12 +\frac{i\mu \beta}{2\pi}\right)\right]\\ \fl
&=&\log{2}+\log{\cosh{\left(\frac{\mu
\beta}{2}\right)}}-\frac{|\mu|\beta}{2}\,.\label{delta2}\eeq

Putting both pieces together, we finally have \beq\nn\log{{\cal
Z}}&=&-\frac{\beta
\pi}{12L}+\sum_{n=1}^{\infty}\log{(1+e^{-\frac{2n\pi
\beta}{L}}+2\cosh{(\mu \beta)}e^{-\frac{n\pi \beta}{L}})}\\
&+&\log{2}+\log{\cosh{\left(\frac{\mu
\beta}{2}\right)}}-\frac{|\mu|\beta}{2}\,.\label{log}\eeq

\subsection{$ \frac{\pi}{L} <|\mu| < \frac{2\pi}{L}$}

Again, we have
\beq \fl
{\zeta}_1(s)=\frac{(1+(-1)^{-s})}{\Gamma(\frac{s}{2})}\int_0^{\infty}dt\,
t^{\frac{s}{2}-1}\!\!\!\!\sum_{\begin{array}{c}
  n=1 \\
  l=-\infty \\\end{array}}^{\infty}\!\!\!\!e^{-t\left[{\left(\frac{n\pi}{\alpha L}\right)}^2+
{\left((2l+1)\frac{\pi}{\alpha
\beta}-i\frac{\mu}{\alpha}\right)}^2\right]}\,.\eeq

However, in this range of $\mu$, the contribution to the zeta
function due to $n=1$ must be analytically extended in a different
way. In fact, the expression cannot be written in terms of a
unique Mellin transform, since its real part is not always
positive (note, in connection with this that, for $n=1$, eq.
(\ref{ext1mu}) diverges). Instead, it can be written as a product
of two Mellin transforms \beq \fl \nn {\zeta}_1^{n=1}
(s)&=&\frac{(1+(-1)^{-s})}{\alpha^{-s}[\Gamma(\frac{s}{2})]^2}\,
 \sum_{l=0 }^{\infty}\int_0^{\infty}dt\,
t^{\frac{s}{2}-1}e^{-\left[ (2l+1)\frac{\pi}{
\beta}-i\mu+i\frac{\pi}{L}\right]t}\\\fl &\times&\int_0^{\infty}dz\,
z^{\frac{s}{2}-1}e^{-\left[ (2l+1)\frac{\pi}{
\beta}-i\mu-i\frac{\pi}{L}\right]z}+\mu \rightarrow -\mu\eeq 
or,
after changing variables according to $t'=t-z; z'=t+z$,  performing the integral over $t'$, and the sum over $l$ \beq \fl \nn
{\zeta}_1^{n=1}
(s)&=&\frac{(1+(-1)^{-s})\sqrt{\pi}}{2\alpha^{-s}\Gamma(\frac{s}{2})}\,
 \left(2\frac{\pi}{L}\right)^{\frac{1-s}{2}}\\\fl
&\times&\int_0^{\infty}dz\, z^{\frac{s-1}{2}}
J_{\frac{s-1}{2}}(\frac{\pi}{L}z)\frac{e^{i\mu z}}{\sinh{(\frac{\pi
z}{\beta})}}+\mu \rightarrow -\mu \,.\eeq

Now, the integral in this expression diverges at $z=0$. In order
to isolate such divergence, we add and subtract the first term in
the series expansion of the Bessel function, thus getting the
following two pieces \beq \fl \nn {\zeta}_{1,(1)}^{n=1}
(s)&=&\frac{(1+(-1)^{-s})\sqrt{\pi}s}{4\alpha^{-s}\Gamma(\frac{s}{2}+1)}\,
 \left(2\frac{\pi}{L}\right)^{\frac{1-s}{2}}\\\fl\nn
&\times&\int_0^{\infty}dz\, z^{\frac{s-1}{2}}
\left[J_{\frac{s-1}{2}}(\frac{\pi}{L}z)- \frac{\left(\frac{\pi z}{2L}\right)^{\frac{s-1}{2}}}
{\Gamma\left(\frac{s+1}{2}\right)}\right]\frac{e^{i\mu
z}}{\sinh{(\frac{\pi z}{\beta})}}\\\fl &+& \mu \rightarrow
-\mu\,,\label{zeta21}\eeq and \beq \fl {\zeta}_{1,(2)}^{n=1}
(s)=\frac{(1+(-1)^{-s})\sqrt{\pi}}{2^s
\alpha^{-s}\Gamma(\frac{s}{2})\Gamma(\frac{s+1}{2})}\, 
\int_0^{\infty}dz\, z^{{s-1}}\frac{e^{i\mu z}}{\sinh{(\frac{\pi
z}{\beta})}}+\mu \rightarrow -\mu \,.\label{zeta22}\eeq

The contribution of equation (\ref{zeta21}) to the partition
function can be easily evaluated by noticing that the factor
multiplying $s$ is finite at $s=0$. Thus, one has
\beq {\Delta}_{1,(1)}^{n=1}=- \int_0^{\infty}\!dz\,z^{-1} \left[\cos{(\frac{\pi}{L}z)}-1\right]\frac{e^{i\mu z}}{\sinh{(\frac{\pi z}{\beta})}}+\mu \rightarrow -\mu \,,\eeq
where we have used that $J_{-\frac{1}{2}}(\frac{\pi}{L}z)=\sqrt{\frac{2}{\pi \frac{\pi}{L}z}}\cos{(\frac{\pi}{L}z)}$. Now, in the term with $\mu\rightarrow -\mu$, one can change $z\rightarrow -z$ to obtain
\beq {\Delta}_{1,(1)}^{n=1}=-\int_{-\infty}^{\infty}dz\,z^{-1} \left[\cos{(\frac{\pi}{L}z)}-1\right]\frac{e^{i\mu z}}{\sinh{(\frac{\pi z}{\beta})}} \,.\eeq
This last integral is easy to evaluate in the complex plane, by carefully taking into account the sign of $\mu$, as well as well as the fact that $\frac{\pi}{L}<|\mu|$ in closing the integration path, to obtain
\beq {\Delta}_{1,(1)}^{n=1}=-2\sum_{l=1}^{\infty}
\left[\frac{(-1)^l}{l}\cosh{(\frac{\pi}{L} \beta l
)}e^{-|\mu|\beta l}+\frac{(-1)^{l+1}}{l}e^{-|\mu|\beta l}\right]\label{serie}\eeq
or, after summing the series 
 \beq \nn {\Delta}_{1,(1)}^{n=1}&=&\left\{\log{\left(1+e^{-2|\mu|\beta}
+2\cosh{(\frac{\pi}{L}\beta)}e^{-|\mu|\beta}\right)}\right.\\
&+&\left.
|\mu|\beta -
2\log\left(2\cosh{(\frac{\mu\beta}{2}})\right)\right\}\,.\label{del21may}\eeq

In order to get the contribution coming from (\ref{zeta22}), the
integral can be evaluated for $\Re s>1$, which gives \beq \nn
{\zeta}_{1,(2)}^{n=1} (s)&=&\frac{(1+(-1)^{-s})\Gamma(s)
\sqrt{\pi}{(\alpha \beta)}^{s}
}{{(2\pi)}^{s}2^{s-1}\Gamma(\frac{s}{2})\Gamma(\frac{s+1}{2})}
\\&\times&\left[\zeta_H (s,\frac12
(1-\frac{i\mu\beta}{\pi}))+\zeta_H (s,\frac12
(1+\frac{i\mu\beta}{\pi}))\right] \,.\eeq  
Its contribution to the partition
function can now be obtained by using that $\zeta_H (0,\frac12
(1-\frac{i\mu\beta}{\pi}))+\zeta_H (0,\frac12
(1+\frac{i\mu\beta}{\pi})=0$ and the well known value of
$-\frac{d}{ds}\rfloor_{s=0}\zeta_H(s,x)$ \cite{gradshteyn}, to
obtain \beq
{\Delta}_{1,(2)}^{n=1}=2\log(2\cosh{(\frac{\mu\beta}{2}}))
\,.\label{del22may}\eeq

Summing up the contributions in equations (\ref{delta2}),
(\ref{del21may}) and (\ref{del22may}), as well as the contribution
coming from $n\geq 2$, evaluated as in the previous subsection,
one gets for the partition function \beq \fl\nn \log{Z}&=&
\left\{\log{\left(2\cosh{\left(\frac{\mu
\beta}{2}\right)}\right)}+\frac{|\mu|\beta}{2}\right.\\ \fl &+&
\nn \log{\left(1+ e^{-2|\mu|\beta}+2\cosh{(\frac{\pi}{L}\beta)}e^{-|\mu|\beta}\right)} +\beta \frac{\pi}{L}\left(-\frac{1}{12}-1\right)\\\fl
&+&\left.\sum_{n=2}^{\infty} \log{\left(1+ e^{-2n\frac{\pi}{L}\beta}+2\cosh{(\mu \beta )}e^{-n
\frac{\pi}{L}\beta}\right)}\right\} \label{zmayor}\,.\eeq

At first sight, this result looks different from the one
corresponding to $|\mu|<\frac{\pi}{L}$ (equation (\ref{log})). However,
it is easy to see that both expressions coincide, the only difference being that the zero-temperature limit is explicitly isolated from
finite-temperature corrections. Similar calculations lead to the same conclusion for other ranges of $\mu$. In those cases where $\mu$ coincides exactly with one energy level, the result can be shown to be the same, but series such as those in equations (\ref{ext1mu}) and (\ref{serie}) are only conditionally convergent.

\section{Free energy and fermion number}\label{Sect4}

From the results in the previous section, the free energy can be readily obtained. It is given by
\beq\fl\nn
F&=&-\frac{1}{\beta}
\log{Z}=\frac{
\pi}{12L}-\frac{1}{\beta}\left[\sum_{n=1}^{\infty}\log{(1+e^{-\frac{2n\pi
\beta}{L}}+2\cosh{(\mu \beta)}e^{-\frac{n\pi \beta}{L}})}\right.\\
\fl&+&\left.\log{2}+\log{\cosh{\left(\frac{\mu
\beta}{2}\right)}}-\frac{|\mu|\beta}{2}\right]\,.\eeq

It is continuous, in particular, at $|\mu| =k \frac{\pi}{L},
 \,k=0,...,\infty$. In the low-temperature limit one has \beq 
F(\frac{k\pi}{L}<|\mu| < \frac{(k+1)\pi}{L})\rightarrow_{\beta\rightarrow
\infty}\frac{\pi}{12L}+k(k+1)\frac{\pi}{2L}-k|\mu|\,.\label{limit}\eeq

The fermion number is obtained as
\beq
N=\frac{1}{\beta}\frac{\partial\,\log{{\cal Z}}}{\partial
\mu}\,.\eeq

It
is given by \beq \nn
N&=&\left\{\sum_{n=1}^{\infty}\left[\frac{e^{-\frac{n\pi
\beta}{L}+\mu \beta}}{1+e^{-\frac{n\pi \beta}{L}+\mu
\beta}}-\frac{e^{-\frac{n\pi \beta}{L}-\mu
\beta}}{1+e^{-\frac{n\pi \beta}{L}-\mu
\beta}}\right]\right.\\&+&\left.
\frac12\tanh\left(\frac{\mu \beta}{2}\right)-\frac12
sign(\mu)\right\}\,,\label{numero} \eeq

Note that it is not defined for $\mu=0$, where both lateral limits differ. This
originates from the indetermination of the phase of the determinant (equation (\ref{ext2mu})). From a physical point of view, this reflects the fact that the sign of $\mu$ distinguishes particles from antiparticles.

Particularly interesting is the discontinuous behavior of $N$ in the zero-temperature limit, where $\mu$ is to be interpreted as a Fermi energy. In such limit, one has
\beq 
N(\frac{k\pi}{L}<|\mu| < \frac{(k+1)\pi}{L})\rightarrow_{\beta\rightarrow
\infty}-k sign(\mu)\,,\eeq
which coincides with the derivative of equation (\ref{limit}), and is consistent with Fermi statistics. For $\mu$ equal to an energy level, both lateral limits differ, and $N(|\mu|=\frac{k\pi}{L})$ is undefined.

\ack{We thank Emilio Elizalde for having made our participation in 
QFEXT05 possible, for his warm hospitality and for the perfect 
organization of the event.

This work was partially supported by Universidad Nacional de 
La Plata (Grant 11/X381).}

\end{document}